\documentclass[12pt,epsfig]{article}
\usepackage{amsmath}
\usepackage{graphicx}
\usepackage{epstopdf}
\usepackage{titlesec}
\usepackage[usenames, dvipsnames]{color}
\titleformat*{\section}{\large\bfseries\sffamily}
\titleformat*{\subsection}{\small\bfseries\sffamily}
\begin{document}
\begin{center}
\textbf {Clustering of Gamma-Ray bursts through kernel principal component analysis}\\
\end{center}
Soumita Modak$^1$, Asis Kumar Chattopadhyay$^1$, and Tanuka Chattopadhyay$^2$\\
$1$: Department of Statistics, Calcutta University, India\\
$2$: Department of Applied Mathematics, Calcutta University, India\\
\begin{abstract}
We consider the problem related to clustering of gamma-ray bursts (from ``BATSE" catalogue) through kernel principal component analysis in which our proposed kernel outperforms results of other competent kernels in terms of clustering accuracy and we obtain three physically interpretable groups of gamma-ray bursts. The effectivity of the suggested kernel in combination with kernel principal component analysis in revealing natural clusters in noisy and nonlinear data while reducing the dimension of the data is also explored in two simulated data sets.
\end{abstract}
keywords:{ clustering; gamma ray bursts; kernel principal component analysis; positive definite kernel.}
\section{Introduction}
Gamma-ray bursts (GRBs), the brightest explosion in the universe since the Big Bang, show huge variation in their duration which can vary from ten milliseconds to several hours, indicating the variation in their formation. To explore the possible sources, clustering of GRBs is performed in different ways (Chattopadhyay et al., 2007 and references therein). Among the controversy that the number of natural groups in GRBs is two or three, we apply kernel principal component analysis (Sch\"{o}lkopf and Smola, 2002) to GRB data set to perform clustering as well as dimension and noise reduction. Previous work of kernel principal component analysis on astronomical data includes classification of supernovae (Ishida et al., 2012, 2013), denoising of early-type galaxies (Modak et al., 2016), etc. \\
Kernel principal component analysis is a nonlinear transformation on raw data, where nonlinear features are extracted from data in terms of kernel principal components. It is a generalization of linear transformation performed in standard principal component analysis, where linear features are extracted from data in terms of principal components. We find that principal component analysis fails to detect the clustering nature in GRB data while kernel principal component analysis finds better results. Clustering based on first few kernel principal components extracted through the proposed kernel, outperforming results of other competent kernels, reveals three physically interpreted groups of $1972$ GRBs.\\
The paper is organized as follows. Section $2$ gives a brief introduction to linear principal component analysis and kernel principal component analysis. In Section $3$, we develop a new kernel, discuss its properties and carry out simulation study. Clustering of GRBs and study variables are discussed in Section $4$. Results with interpretation are given in Section $5$ and Section $6$ concludes.
\section{Principal component analysis and kernel principal component analysis}
\subsection{Principal component analysis (PCA)}
PCA extracts structure from data in terms of resulted features called principal components (PCs), which are ordinarily obtained from orthogonal transformation of the data set.
Given a set of M centered observations $x_{k}, k=1, 2, ..., M, x_{k}\in R^{N}, \sum_{k=1}^{M} x_{k}=0$, PCA diagonalizes the covariance matrix (for simplification, sample covariance matrix is used in place of population version)
\begin{equation}
C=\frac{1}{M} \sum_{j=1}^{M}x_{j}x_{j}^{T},
\end{equation}
in which PCs are obtained by solving the equation
\begin{equation}
\lambda V=CV,
\end{equation}
for eigenvalues $\lambda\geq0$ (as $C$ is non-negative definite matrix), and eigenvectors $V\in R^{N}$ (non-zero vector).
Let $\lambda_{1}\geq...\geq\lambda_{M}\geq0$ be the eigenvalues of $C$ with $\lambda_{l}$ being the last nonzero eigenvalue and $ V^{1}, ..., V^{M}$ be the corresponding orthonormal set of eigenvectors. Then for given random vector $x\in R^{N}$, the $k^{th}$ principal component of $x$ is obtained by
\begin{equation}
u_{k}={V^{k}}\cdot x, \hspace{1 mm}k=1,2,...,l.
\end{equation}
For nonlinear data structure, PCA, which is based on linear transformation, must fail. PCs also extract noise from a noisy data. To overcome such drawbacks of PCs, we consider the methods of analysis described in the following subsections.
\subsection{Kernel principal component analysis (KPCA)}
Kernel principal component analysis (Sch\"{o}lkopf and Smola, 2002), a generalization of principal component analysis, is performed on a dot product space F (feature space) instead of the input space $R^{N}$ by using a map from $R^{N}$ to F
\begin{equation*}
\Phi:R^{N}\rightarrow F.
\end{equation*}
Then for $\sum_{k=1}^{M} \Phi(x_{k}) = 0$, the covariance matrix $\bar{C}$ in F can be written as,
\begin{equation}
\bar{C}=\frac{1}{M} \sum_{j=1}^{M}\Phi(x_{j}) \Phi(x_{j})^{T}.
\end{equation}
Now, we need to solve the following equation for eigenvalues $\lambda\geq0$, and eigenvectors $V\in F$ (non-zero vector).
\begin{equation}
\lambda V=\bar{C}V.
\end{equation}
\subsection{Kernel trick}
For some coefficients $\alpha_{i}\in R$ $(i=1, 2, ..., M)$ $V$ can be written as
\begin{equation*}
V=\sum_{i=1}^{M} \alpha_{i} \Phi(x_{i}),
\end{equation*}
which gives rise to an $M\times M$ symmetric, positive semi-definite matrix $K=((K_{ij}))_{i,j=1(1)M}$, given by
\begin{equation}\label{kt}
K_{ij}=<\Phi(x_{i}), \Phi(x_{j})>,
\end{equation}
where $<\cdot,\cdot>$ represents the usual dot product.
By the above kernel trick we avoid computing the map $\Phi$ in F, which can have arbitrarily large dimension. Instead, we simply compute the dot product in F, which leads to just solving the eigenvalue problem for kernel matrix $K=((K_{ij}))_{i,j=1(1)M}$ as follows
\begin{equation}
M \lambda \alpha=K \alpha.
\end{equation}
Any positive definite kernel can be used in (\ref{kt}), where F is a reproducing kernel Hilbert space satisfying $\forall$ $\Phi\in$F, $||\Phi||=\sqrt{<\Phi,\Phi>}$ (Hofmann et al., $2008$).\\
\subsection{Nonlinear or kernel principal components}
Let $\lambda_{1}\geq\lambda_{2}\geq...\geq\lambda_{M}\geq0$ denote the eigenvalues of $K$ and $\alpha^{1},...,\alpha^{M}$ be the corresponding eigenvectors, where $\lambda_{l}$ is the last nonzero eigenvalue. Normalization of eigenvectors $V^{1},...,V^{l}$ in F can be written in terms
of $\alpha^{1},...,\alpha^{l}$ as
\begin{equation*}
\alpha^{k}\cdot\alpha^{k}=1/\lambda_{k}, \hspace{1 mm}k=1,2,...,l.
\end{equation*}
Then the $k^{th}$ kernel principal component (KPC) corresponding to $\Phi(x)$ (Sch\"{o}lkopf and Smola, 2002) is given by
\begin{equation}
<V^{k},\Phi(x)>=\sum_{i=1}^{M} \alpha^{k}_{i} <\Phi(x_{i}),\Phi(x)>=\sum_{i=1}^{M} \alpha^{k}_{i} k(x_{i},x), \hspace{1 mm}k=1,2,...,l,
\end{equation}
where $ k(x_{i},x)$= kernel corresponding to $x_{i}$ and $x$.\\
Next we relax the assumption of being centered on the observations. In that case, kernel matrix $\tilde{K}$ is used instead of $K$, where
\begin{equation}
\tilde{K}_{ij}=(K-1_{M}K-K1_{M}+1_{M}K1_{M})_{ij},\hspace{1 mm}(1_{M})_{ij}=1/M, \hspace{1 mm} for \hspace{1 mm} i,j=1,2,...,M.
\end{equation}
This enables KPCA to be performed using conditionally positive definite kernels $\supset$ positive definite kernels (Hofmann et al., $2008$).\\
No theoretical result exists on what number of KPCs is to be extracted to gain sufficient information on data under study. So, first we extract the first KPC and analysis is performed on that. Then we do the same based on the first two KPCs and so on. We continue in this way until we obtain improvement in terms of some accuracy measure chosen appropriately in the context of study (e.g. classification error in Sch\"{o}lkopf and Smola, 2002). KPCA is performed on GRB data set for nonlinear feature extraction and dimension reduction. Noise is simultaneously and automatically reduced by discarding projections of transformed data onto the higher order eigenvectors in feature space. Thus we have found the natural clustering nature in GRBs.
\section{Proposed kernel}
In this paper we propose the real-valued symmetric kernel
\begin{equation} \label{my kernel}
k(x,y)=\exp(- {\sum\limits_{i=1}^N |\frac{x_{i}-y_{i}}{s_{i}}|^p}),
\end{equation}
between $x,y\in R^{N}$, where $p$ $(0<p\leq 2)$ is a tuning parameter and $s_{i}$ $(>0)$$, i=1, 2,..., N$ are scale parameters, called hyperparameters. Here it can be seen that, as $x\leftrightarrow y, \hspace{1 mm} k(x,y)\downarrow 0$. Moreover, $k(x,y)\in(0,1]$ and is positive definite (p.d.) when $p\in(0,2]$, since $\exp(-|x-y|^{p}), x,y\in R$  is p.d. when $ 0<p\leq 2$ (Berg et al., 1984) and equation (\ref{my kernel}) is the dot product of $N$ such p.d. kernels (Hofmann et al., 2008). In the kernel (\ref{my kernel}), we can fix the problem of divergence by the parameter $p$ when two points in $R^{N}$ are far away. For example, by choosing a small $p$ we can prevent the numerical divergence, which often occurs in radial basis function, provided the exponent should not be zero.\\
Next, we compare the kernel (\ref{my kernel}) with largely used kernels existing in the literature. Some possible choices of $k(x,y)$ are given below.\\
(i) Polynomial kernel of degree d:
\begin{equation} \label{poly}
k(x,y)=(<x,y>+c)^d,
\end{equation}
where $c\geq0$ and $d$ is a positive integer. Here, in particular, when $c=0$ and $d=1$, kernel PCA becomes standard PCA.\\
(ii) Gaussian radial basis function kernel:
\begin{equation}  \label{gaussian}
k(x,y)=\exp(- \frac{\parallel x-y\parallel^{2}}{2\sigma^{2}}),
\end{equation}
where $\sigma > 0$ is a scale parameter and $\parallel\cdot\parallel=$Euclidean norm. For $p=2$ and $s_{i}=\sqrt2\sigma$ for all $i=1,2,...,N$, kernel (\ref{my kernel}) boils down to kernel (\ref{gaussian}).\\
(iii) Laplacian kernel:
\begin{equation} \label{lap}
k(x,y)=\exp(- \frac{\parallel x-y\parallel}{\sigma}).
\end{equation}
For $p=1$ and $s_{i}=\sigma$ for all $i=1,2,...,N$, kernel (\ref{my kernel}) boils down to kernel (\ref{lap}).\\
(iv) Sigmoid kernel:
\begin{equation} \label{sig}
k(x,y)=tanh(a<x,y>+b),
\end{equation}
where $a,b\geq0$.\\
\subsection{Effectiveness of proposed kernel through simulation study}
For graphical representation, we consider only two dimensional data sets. First, we draw random vectors of size 500 from bivariate entangled spiral which has two natural groups, known a priori. Then Gaussian noise with standard deviation 0.05 is added to the data. Fig.1 shows the first KPC, extracted through kernel (\ref{my kernel}) with $p=1/2, s_{1}=0.07, s_{2}= 0.13 $, is sufficient to reveal the two groups in the noisy data and hence reduces the dimension of the data. Next, random vectors of size 500 are obtained from nonlinear data sets in which four differently shaped distributions (viz. Gaussian, square, triangle and wave) are used. The first two KPCs, extracted through kernel (\ref{my kernel}) with $p=1/2, s_{1}=1.24, s_{2}=  1.89$, expose all the four heterogeneous groups present in the data (Fig.2).
\section{Clustering of GRBs}
Our data set, retrieved from the fourth BATSE Gamma-Ray Burst Catalog (revised) (Paciesas et al., 1999), consists of information on $1972$ GRBs for the following $9$ variables. $F_1,F_2,F_3,F_4$ are time-integrated fluences in  $20-50$, $50-100$, $100-300$ and $>300$ keV spectral channels respectively; $P_{64},P_{256},P_{1024}$ are peak fluxes measured in $64, 256$ and $1024$ ms bins respectively; $T_{50},T_{90}$ are times within which $50\%$ and $90\%$ of the flux arrive. Unit of fluence is given in ergs per square centimetre (ergs cm$^{-2}$), unit of peak flux is count per square centimetre per second (cm$^{-2}$ s$^{-1}$) and unit of time is second (s).\\
First, observations on each variable are standardized because the ranges of the variables vary largely (Table \ref{table:t1}). Then, for a particular choice of kernel, KPCA is performed on them. We extract nonlinear features using the significant kernel principal components. Then using them as study variables, k-means clustering method (Hartigan--Wong clustering algorithm; Hartigan et al., 1979) is performed on them in which the number of clusters is determined with the help of gap statistic (Tibshirani et al., 2001).\\
\subsection{Choice of kernel principal components and hyperparameters}
KPCs are supposed to carry less information and more noise with increasing order and after a certain order they fail to give any relevant information regarding the data under study. So, we start by choosing the first few KPCs and the number of chosen KPCs is increased as long as their performance gets better in terms of an accuracy measure. In this context, we choose the Dunn index (Dunn, 1974) as our accuracy measure, which indicates the internal validation of a clustering performed. It takes value between $0$ and $\infty$ with greater value indicating better clustering.\\
In kernel (\ref{my kernel}), a plausible choice for hyperparameter $s_{i}$ is the square root of the sample variance of the $i^{th}$ variable, $i=1, 2,...,N$. As the analysis is performed based on the data standardized for each variable, we consider $s_{1}=s_{2}=...=s_{N}=s$ (say) with a plausible choice for $s$ being the square root of the sample variance of the whole data set. 95\% bootstrap confidence interval (the non-parametric BCa interval, Efron and Tibshirani, 1993) for $s$ is computed in which lower BCa limit, upper BCa limit and thier arithmetic mean are represented by $\sigma_{1},\sigma_{2}$ and $\sigma_{3}$ respectively. Next, for the pair of hyperparameters  $s$ and $p$, grid search method is applied to the set $\{\sigma_{1}, \sigma_{2}, \sigma_{3}\}\times\{2, 1, 1/1.5, 1/2\}$. For comparison purpose, we consider the same values for $\sigma$ in kernel (\ref{gaussian}) and  kernel (\ref{lap}) as for $s$ in kernel (\ref{my kernel}).\\
Now kernel principal component analysis is functioned for a particular choice of kernel with corresponding hyperparameters. For each value of hyperparameter (or each set of values for hyperparameters) considered, KPCA is performed independently. At first, k-means clustering is acted on the basis of the first kernel principal component in which number of clusters is determined by gap statistic. Then based on the obtained clusters we compute Dunn index. Next, we take the first two kernel principal components and do the same. Like this we gradually go up to higher order kernel principal components to extract more information till improvement is met in terms of Dunn index. Maximum Dunn index giving KPC (or set of KPCs) is chosen under a fixed choice of values for hyperparameters. We repeat this algorithm for each choice of values for hyperparameters and finally maximum Dunn index giving combination of values for hyperparameters and KPCs is selected.
\subsection{Robustness of KPCA w.r.t. clustering method and accuracy measure}
Here clustering of GRBs is performed based on KPCs extracted through KPCA in association with k-means clustering method where Dunn index is used as accuracy measure. Now for fixed choice of KPCs we can vary the clustering method and accuracy measure. When the extracted components are able to show up the natural clustering in GRBs then use of other clustering methods and accuracy measures should give the same results with three physically interpretable groups of GRBs. To show the robustness with respect to accuracy measure we compute $11-$ NN ($11-$nearest neighbor) classification error rate (Ripley, 1996) for KPCs with the highest Dunn index. While to show the robustness with respect to clustering method we perform hierarchical cluster analysis using average linkage (Kaufman and Rousseeuw, 1990) on Euclidean distance matrix computed on KPCs with the highest Dunn index. Here the number of clusters is chosen by the average Silhouette width (ASW) (Rousseeuw, 1987). ASW ($-1\leq$ ASW $\leq 1$) with a large value (close to 1) indicates very well clustering, having small value (around 0) means the observations lie between two clusters, and ASW with a negative value indicates that observations are probably placed in the wrong clusters.
\section{Results and interpretation}
First, k-means clustering method is applied to the standardized variables of GRB data set in which gap statistic indicates no clustering present in GRBs, i.e. raw GRB data set fails to reveal the inherent clustering nature in GRBs. Then the same method is applied to the principal components, extracted from the GRB data through principal component analysis. Linear features (first two PCs extracted through kernel $(11)$ with $c=0,d=1$), explaining more than $80\%$ variation in data, result in one group of GRBs. Thus linear information on data also can not expose the natural groups (two or three) present in GRBs (Table \ref{comparison}).\\
Kernel (\ref{my kernel}) successfully reveals the inherent clustering nature in GRBs, by extracting the relevant nonlinear information from raw data in terms of kernel principal components. Table \ref{comparison} shows accuracy measure for clustering based on KPCs, extracted through different kernels with different choices of hyperparameters. We see the first two KPCs, extracted through kernel (\ref{my kernel}) with $p<1$ and for every choice of $s$ considered, are enough to describe the data. They sufficiently extract the important nonlinear features from the noisy raw data and hence effectively reveal the three physically interpretable clusters in GRBs. While inclusion of the third KPC results in either no clustering or worse clustering in terms of our chosen accuracy measure. This indicates that the third KPC bears mostly noise and does not account for any usable information in the data set. Hence the third KPC is excluded and no higher order KPCs are considered further. Maximum Dunn index giving combination of hyperparameters is kernel (\ref{my kernel}) with $p=1/2$  and $s=\sigma_{1}$, which outperforms all the other kernels in terms of Dunn index with a value $0.018853$. Table \ref{difam} giving $11-$NN classification error rate for the KPCs with the highest Dunn index value shows the robustness of method of KPCA in clustering GRBs w.r.t accuracy measure. The corresponding value for the first two KPCs, extracted through kernel (\ref{my kernel}) with $p=1/2$  and $s=\sigma_{1}$ is $0.15 \%$ which is quite a favorable value in its selection.\\
We compare our result with clustering of 1594 GRBs performed in Chattopadhyay et al. (2007), in terms of $1$-NN classification error rate (using leave-one-out cross validation on the whole data set on which clustering is performed; Ripley, 1996). In Chattopadhyay et al. (2007), k-means clustering approach is directly applied to differently chosen study variables and 1594 GRBs are clustered in three groups of sizes 622, 423, and 549 respectively with 4.08 \% $1$-NN classification error rate. While our clustering of 1972 GRBs based on the first two kernel principal components, extracted by kernel (\ref{my kernel}) with $p=1/2$  and $s=\sigma_{1}$, groups those 1594 GRBs in three clusters of sizes  827, 438, and 329 respectively with $0.2\%$ $1$-NN classification error rate.\\
k-means applied to the first two KPCs, extracted through kernel (\ref{my kernel}) with $p=1/2$  and $s=\sigma_{1}$, clusters 1972 GRBs into three groups, say cluster I (k$_1$), cluster II (k$_2$) and cluster III (k$_3$) (Fig.3), in terms of gap statistic. Table \ref{asw} shows that hierarchical clustering method also gives three clusters of GRBs (corresponding dendrogram shown in Fig.\ref{den}) with ASW$=0.64$ (for cluster-wise Silhouette plot see Fig.\ref{silplot}), which is quite a high value evidencing well clustering. We compare the properties based on cluster averages obtained from hierarchical clustering (Table \ref{cluspr1}) and k-means clustering (Table \ref{cluspr}). Similarities between $k_1$ and group$_1$, $k_2$ and group$_3$, $k_3$ and group$_2$ show the robustness of KPCA with respect to clustering method in revealing the three natural clusters present in GRBs.\\
Further discussion on astrophysical properties of the groups is performed based on the results of k-means method. To explore the physical interpretation of the clusters, time ($log_{10}(T_{90})$ in s) vs. fluence ($log_{10}(F_T)$ in ergs $cm^{-2}$, where $F_T=F_1+F_2+F_3+F_4$) graph is shown in Fig.\ref{TimevsFluencewithlines}. k$_3$ is consistent with the short duration bursts (viz. $T_{90}=1.16 \pm  0.07$ s) separated by the solid line $F_T=10^{-5.4}/T_{90}^{0.9}$ (Fig.\ref{TimevsFluencewithlines}). This cluster can be connected with mergers of neutron star systems. The standard long duration bursts ($T_{90}>2$ s) are clustered into two groups, I and II respectively. The solid line $F_T=10^{-4.6}/T_{90}^{0.4}$ separates these two groups. Fig.\ref{HistFt} and Fig.\ref{HistT90} show the histograms for $log_{10}(F_T)$ and $log_{10}(T_{90})$ respectively. As we can see from Fig.\ref{HistT90} that the distribution of $log_{10}(T_{90})$ for GRBs is bimodal based on which Kouveliotou et al. (1993) separates GRBs into two classes as short events ($< 2$ s) and longer events ($> 2$ s). But multivariate analysis considering other important variables rather than only the duration clusters GRBs into three physically interpretable groups (Balastegui et. al, 2001; Chattopadhyay et al., 2007; Veres et al., 2010) which is consistent with our findings corresponding to short duration bursts ($k_{3}$), intermediate duration bursts ($k_{2}$), and long duration bursts with higher fluence ($k_{1}$). We also plot the diagram of $log_{10}(T_{90})$ vs. $log_{10}(H_{32})$ (Fig.\ref{H32}), where $H_{32}=F_3/F_2$ is a measure of spectral hardness. We can see that the short duration bursts having the highest average hardness ratio (viz. $H_{32}=6.17\pm0.25$) is consistent with that of Veres et al. (2010). Veres et al. (2010) performing the cluster analysis of $408$ GRBs, collected from a different data source, on the basis of $H_{32}$ and $T_{90}$ says that the group of intermediate duration bursts is the softest one. But the intermediate and long duration bursts in our study have more or less similar average hardness ratios (viz. $H_{32}=3.25\pm0.12$ and $H_{32}=3.29\pm0.06$ for $k_2$ and $k_1$ respectively). Thus unlike the previous authors we cannot say that they are X-Ray flashes (XRF) and X-Ray rich (XRR) populations. Also in their work the observed XRFs do not cover all the intermediate GRBs. Therefore we can physically interpret them  in the following manner. \\
In a previous work, Ghirlanda et al. (2004) measures redshift corrected intrinsic energy spectrum, E$_{peak}$ (in keV) and isotropic energy output, E$_{iso}$ (in ergs) for 25 GRBs in which E$_{peak}$ and E$_{iso}$ are found correlated. In order to see how our work affects this relationship, we take these GRBs which have measured values of F$_T$ and T$_{90}$ as well. On the basis of the separating lines between the groups we separate these GRBs and plot E$_{peak}$ vs. E$_{iso}$ in Fig.\ref{ClassificationGRB}. It is clear from Fig.\ref{ClassificationGRB} that low fluence GRBs mostly fall in k$ _{2}$ and they have more or less constancy in isotropic energy output ( viz. E$_{iso}$). So this group might be associated with the neutron star-white dwarf mergers (King et al., 2007), since both of neutron stars and white dwarfs do not have significant mass variations leading to nearly constant energy output. Also, their merger time is smaller which is consistent with shorter intrinsic duration of T$_{90}$ (viz. $T_{90}=18.53 \pm 0.59$ s). On the other hand, massive stellar collapse might be associated with the members of  k$_1$. Because energy output and duration of a GRB connected with massive stellar collapse depend on the mass and size of the progenitor, which is observed in the variation inferred for high fluence burst. The interpretation of these two groups will be more convincing subject to future observations of higher fluence GRBs having highest energy outputs (viz. $~10^{52}-10^{54}$ ergs).\\
\section{Conclusion}
This work concerns clustering of gamma-ray burts. Here we not only reduce the burden of the data, but also extract the inherent information from the data, on which simple clustering method reveals the natural groups in GRBs. We propose a new possible way, kernel principal component analysis, to analyze GRB data set as well as a new kernel, which makes the clustering results better in comparison with the other existing kernels and gives three physically interpretable groups in GRBs. However explanation of the sources of these three groups will be more prominent in the future with more data collection.
\clearpage
\begin{table}
\caption{Variable-wise range of GRB data set}
\begin{center}
\begin{tabular}{c c }
\hline
Variable & Range\\
\hline
$F_1$ & (0, 4.61e-05)\\
$F_2$ & (0, 3.72e-05) \\
$F_3$ & (0, 0.000139)\\
$F_4$ & (0, 0.000608)\\
$P_{64}$ &  (0, 183.37)\\
$P_{256}$ & (0, 181.634)\\
$P_{1024}$ &(0, 163.344)\\
$T_{50}$ & (0.012, 481.984)\\
$T_{90}$   & (0.024, 673.824) \\
\hline
\end{tabular}
\end{center}
\label{table:t1}
\end{table}
\clearpage
\begin{table}
\caption{Comparison of different kernels}
\begin{center}
\tiny
\begin{tabular}{c c c c c c}
\hline
Kernel & Hyperparameter(s) & No. of  & No. of & Cluster size(s)  & Dunn Index \\
       &                 & first KPC(s) & cluster(s)     &        &  ($\times 10^{4}$)\\
\hline

kernel(\ref{my kernel})  & $s=\sigma_{1},p=2^{\P}$       &- &-&-& -\\

                         & $s=\sigma_{2},p=2^{\P}$     &- &-&-&- \\
                         & $s=\sigma_{3},p=2^{\P}$  &- &-&-&- \\

kernel(\ref{my kernel})  & $s=\sigma_{1},p=1$      & 1               & 2               &  1441,  531               & 84.47      \\

                         &                       & 2               & 4              &  1030,  165,  370,  407   &  63.49     \\
                         & $s=\sigma_{2},p=1$       & 1               & 2               & 1441,  531          & 84.47     \\
                        &                       & 2               & 4                & 1030,  165,  370,  407   &    63.49           \\

                         & $s=\sigma_{3},p=1$       &  1              & 2              &  1494,  478           &154.03\\
                         &                       & 2               & 4              & 1053,  154,  381,  384            &60.51\\
kernel(\ref{my kernel})  & $s=\sigma_{1},p=1/1.5$    & 1               & 3               & 301, 694, 977       &50.52\\

                         &                       & 2               &3                &  758, 705, 509               & 68.81 \\
                         &                       & 3               & no clustering    &  -                 &   -       \\

                         & $s=\sigma_{2},p=1/1.5$   & 1               & 3                &  303,  662, 1007             &44.71  \\
                         &                       & 2               & 3                &  726 758 488                &  106.15      \\
                         &                       & 3               & no clustering    &  -                 &   -       \\
                         & $s=\sigma_{3},p=1/1.5$   & 1               & 3                 &  282,  634, 1056                 & 73.09  \\   \\
                         &                       & 2               & 3                & 702, 789, 481        &   75.22  \\
                         &                       & 3               & no clustering    &  -                 &   -   \\

kernel(\ref{my kernel})  & $s=\sigma_{1},p=1/2$     & 1               & 5               & 277, 193, 412, 845, 245       &27.69\\

                         &                          & 2               &3                &  941, 588, 443                & 188.53 \\

                         &                       & 3               &5                &  296, 284, 313, 750, 329      & 119.75    \\
                         & $s=\sigma_{2},p=1/2$     & 1               & 3                &  984, 415, 573               &73.38  \\
                         &                       & 2               & 3                &  915, 611, 446                &   88.69       \\
                         &                       & 3               & no clustering    &  -                 &   -       \\
                         & $s=\sigma_{3},p=1/2$     & 1                & no clustering    &  -                 &   -   \\   \\
                         &                       & 2               & 3                & 886, 621, 465        &   60.43  \\
                         &                       & 3               & no clustering    &  -                 &   -   \\

kernel(\ref{poly})       &   $c=0, d=1$          &  1              &   1              &    1972             &-\\
                         &                       &  2              &   1              &    1972              &-\\
                         &                       &  3              &   5              & 80, 318, 1545, 7, 22  & 14.5     \\
                         &                       & 4               & no clustering    &  -                 &   -   \\
                         & $c=0, d=2$            &  1              &   1              &    1972             &-\\
                         &                       &  2              &   2              &  7, 1965            &-\\
                         &                       &  3              &   2              &  7, 1965            &-\\
kernel(\ref{gaussian})  & $\sigma=\sigma_{1}^{\P}$ &- &-&-&-\\
                        & $\sigma=\sigma_{2}^ {\P}$ &- &-&-&-\\
                        & $\sigma=\sigma_{3}^ {\P}$    &- &-&-&-\\

kernel(\ref{lap})      & $\sigma=\sigma_{1}$     & 1               & 3                &439, 789, 744    &  42.25    \\
                       &                         & 2               & 3                &653, 532, 787        &  113.46   \\
                       &                         & 3               & 5                & 275, 278, 585, 483, 351  & 162.48     \\
                       &                         & 4               & no clustering    &  -                 &   - \\
                       & $\sigma=\sigma_{2}$     & 1               & 3                & 442, 774, 756            &34.17    \\
                       &                         & 2               & 3                &638, 539, 795             &69.89  \\
                       &                         & 3               & 5                & 278, 282, 593, 466, 353  &108.78  \\
                       &                         & 4               & no clustering    &  -                 &   -   \\
                       & $\sigma=\sigma_{3}$     & 1               & 5                &295, 575, 580, 256, 266   &19.62  \\
                       &                         & 2               & 3                & 622, 544, 806            & 59.73\\
                       &                         & 3               & no clustering    &  -                 &   -   \\
kernel(\ref{sig})      & $a=1, b=0$              &  1              &   3              &   390,  415, 1167   & 27.65 \\
                       &                         &  2              &   3              &  279,  452, 1241      & 54.85   \\
                       &                         & 3               & 5                & 171,  294, 1074, 204, 229 & 150.18 \\
                       &                         & 4               & no clustering    &  -                 &   -  \\
\hline
$\P$ exponent diverges\\
$BCa_{l}=0.94$,\\
$BCa_{u}= 1.08$\\
\end{tabular}
\end{center}
\label{comparison}
\end{table}
\clearpage
\begin{table}
\caption{Number of clusters w.r.t. classification error rate}
\begin{center}
\tiny
\begin{tabular}{c c c c c c c}
\hline
Kernel & Hyperparameters & No. of  & No. of & Cluster sizes  & Dunn Index               &$11-$NN classification \\
       &                 & first KPC(s) & clusters     &        &  ($\times 10^{4}$)      &error rate ($\times 10^{4}$)\\
\hline

kernel(\ref{my kernel})  & $s=\sigma_{1},p=1/2$     & 1               & 5               & 277, 193, 412, 845, 245       &27.69       &30.43\\

                         &                          & 2               &3                &  941, 588, 443                & 188.53     &15.21\\

                         &                          & 3               &5                &  296, 284, 313, 750, 329      & 119.75     &65.92\\

\hline
\end{tabular}
\end{center}
\label{difam}
\end{table}
\clearpage
\begin{table}
\caption{ASW for hierarchical clustering}
\begin{center}
\begin{tabular}{c c }
\hline
No. of & ASW\\
clusters & (x $10^{2}$)\\
\hline
2 & 56.57\\
3 & 63.58\\
4 & 58.61\\
5 & 60.01\\
\hline
\end{tabular}
\end{center}
\label{asw}
\end{table}
\clearpage
\begin{table}
\caption{Properties of three groups from hierarchical clustering}
\begin{center}
\tiny
\begin{tabular}{c c c c c c c c}
\hline
Name of     &  Size of     & $F_{T} \times 10^{6}$ & $T_{90}$  & $T_{50}$	& $P_{64}$	& $P_{256}$	& $P_{1024}$\\
the cluster & the cluster  & (ergs cm$^{-2}$)     &  (s) & (s)   & (cm$^{-2}$ s$^{-1}$) & (cm$^{-2}$ s$^{-1}$) & (cm$^{-2}$ s$^{-1}$)\\
\hline
group$_1$        & 1079  & 22.14 $\pm$ 1.73 &  62.94$\pm$ 2.20 & 27.32 $\pm$ 1.37  &  5.86 $\pm$ 0.37  & 4.96 $\pm$ 0.31  &  3.69 $\pm$ 0.24\\
group$_2$       & 397   &  0.45 $\pm$ 0.02  & 0.94$\pm$ 0.06  &  0.37 $\pm$ 0.02  &  1.83 $\pm$ 0.05  & 1.07 $\pm$ 0.03  &  0.41 $\pm$ 0.01\\
group$_3$        & 496   &  1.56 $\pm$ 0.06  & 14.37$\pm$0.53  & 5.14 $\pm$ 0.20   &  1.64 $\pm$ 0.07  & 1.17 $\pm$ 0.04  &  0.72 $\pm$ 0.02\\
\hline
\end{tabular}
\end{center}
\label{cluspr1}
\end{table}

\clearpage
\begin{table}
\caption{Properties of three groups k$_1$, k$_2$ \& k$_3$ from k-means clustering}
\begin{center}
\tiny
\begin{tabular}{c c c c c c c c}
\hline
Name of     &  Size of     & $F_{T} \times 10^{6}$ & $T_{90}$  & $T_{50}$	& $P_{64}$	& $P_{256}$	& $P_{1024}$\\
the cluster & the cluster  & (ergs cm$^{-2}$)     &  (s) & (s)   & (cm$^{-2}$ s$^{-1}$) & (cm$^{-2}$ s$^{-1}$) & (cm$^{-2}$ s$^{-1}$)\\
\hline
$k_1$        & 941   & 24.95 $\pm$ 1.96 &  68.02 $\pm$ 2.47 & 29.65 $\pm$ 1.56  &  6.48 $\pm$ 0.42  & 5.49 $\pm$ 0.35  &  4.09 $\pm$ 0.27\\
$k_2$        & 588   &  1.96 $\pm$ 0.07  & 18.53 $\pm$ 0.59 &  6.94 $\pm$ 0.25  &  1.59 $\pm$  0.06 & 1.19 $\pm$ 0.04  &  0.79 $\pm$ 0.02\\
$k_3$        & 443   &  0.49 $\pm$ 0.02  & 1.16 $\pm$  0.07 & 0.45 $\pm$ 0.03   &  1.87 $\pm$ 0.05  & 1.09 $\pm$ 0.03  &  0.43 $\pm$ 0.01\\
\hline
\end{tabular}
\end{center}
\label{cluspr}
\end{table}
\clearpage

\begin{figure}
\centering
\includegraphics[width=1\textwidth]{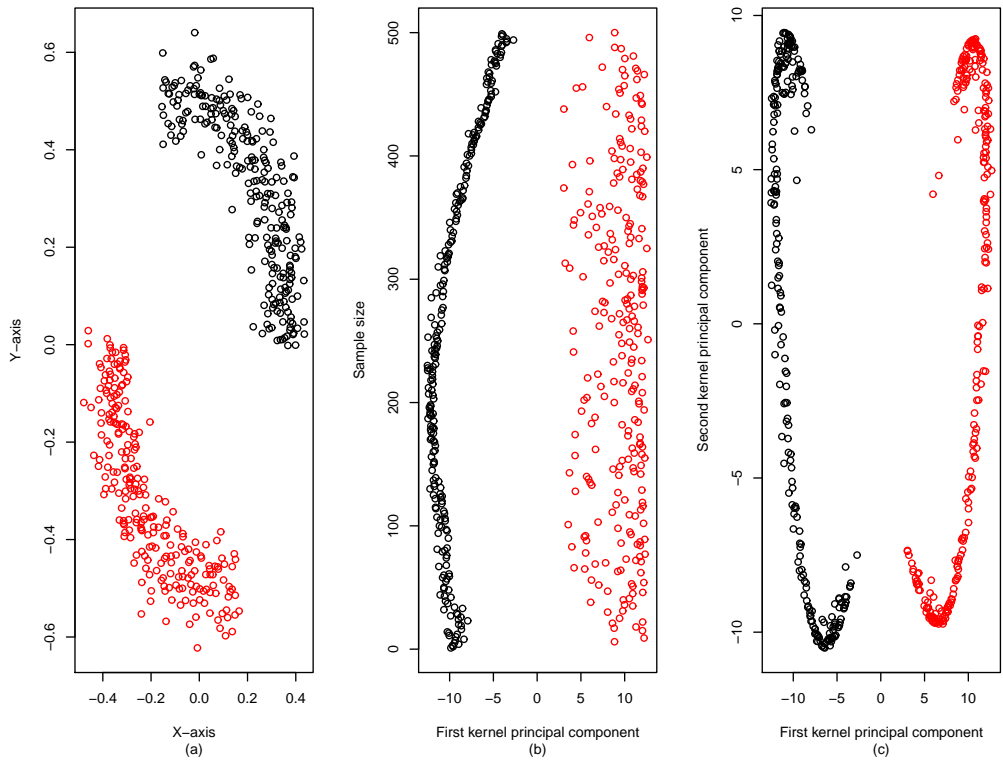}
\caption{(a) Random vectors of size 500 generated from bivariate entangled spiral which has two natural groups (known a priori), with added Gaussian noise having standard deviation$=0.05$, (b) First kernel principal component, extracted through kernel (\ref{my kernel}) with $p=1/2, s_{1}=0.07, s_{2}= 0.13 $, sufficiently revealing the two inherent groups in the noisy data, (c) First two kernel principal components doing the same with higher discrimination.}\label{spiral}
\end{figure}
\clearpage
\begin{figure}
\centering
\includegraphics[width=1\textwidth]{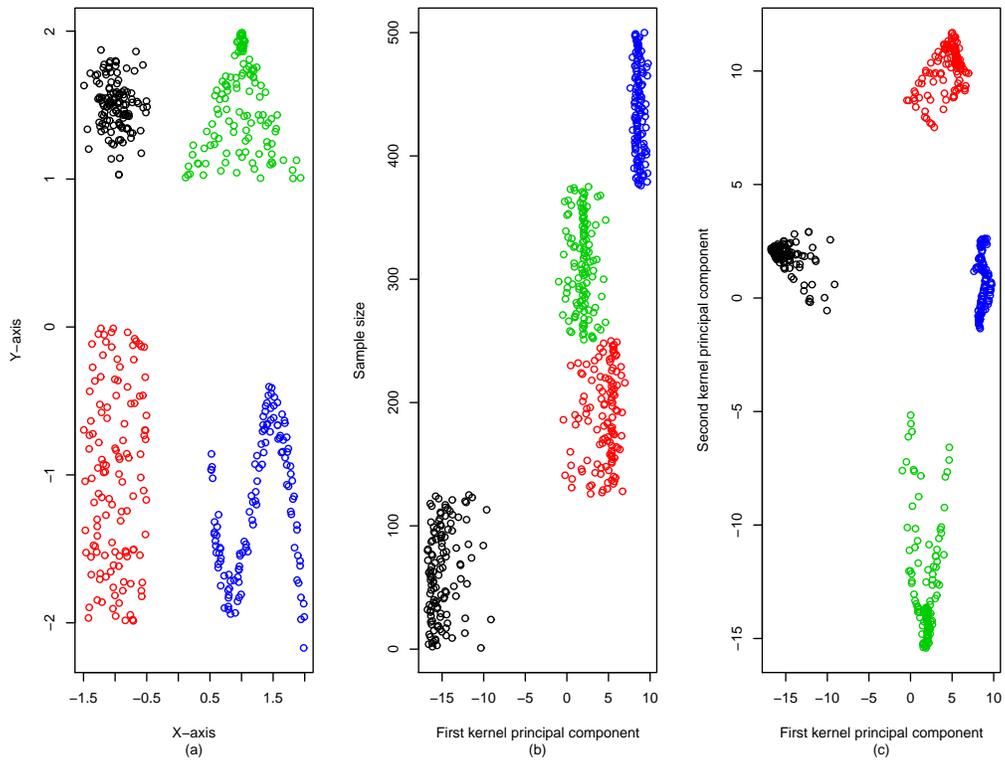}
\caption{(a) Random vectors of size 500 generated from four differently shaped distributions: Gaussian, square, triangle and wave, (b) First kernel principal component, extracted through kernel (\ref{my kernel}) with $p=1/2, s_{1}=1.24
, s_{2}=  1.89$, revealing three inherent groups in data of heterogeneous shapes, while (c) First two kernel principal components successfully revealed all the four heterogeneous groups present in the data.}\label{shapes}
\end{figure}
\clearpage
\begin{figure}
\centering
\includegraphics[width=1\textwidth]{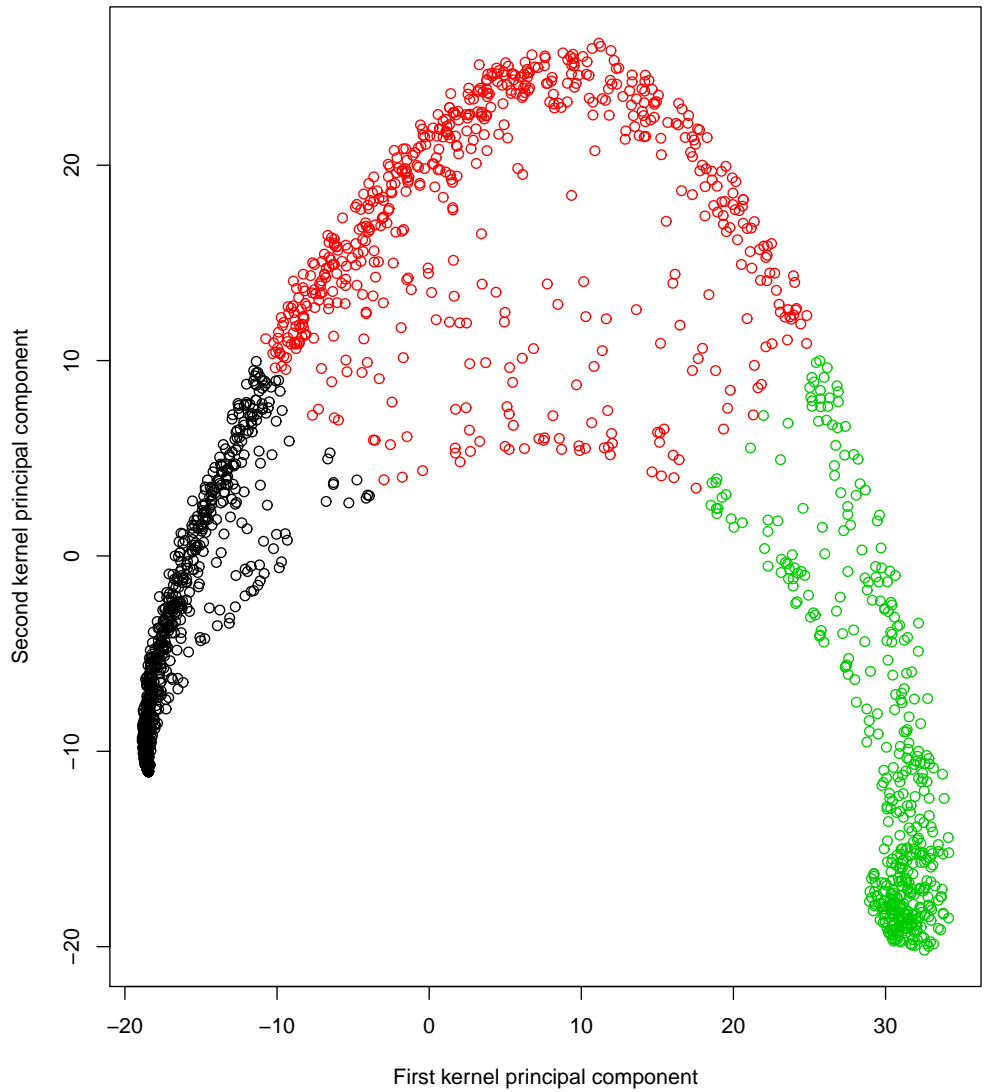}
\caption{K-means applied to the first two KPCs, extracted through kernel (\ref{my kernel}) with $p=1/2$ and $s=\sigma_{1}$, revealed three clusters of GRBs.}\label{ClusterPCs}
\end{figure}
\clearpage
\begin{figure}
\centering
\includegraphics[width=1\textwidth]{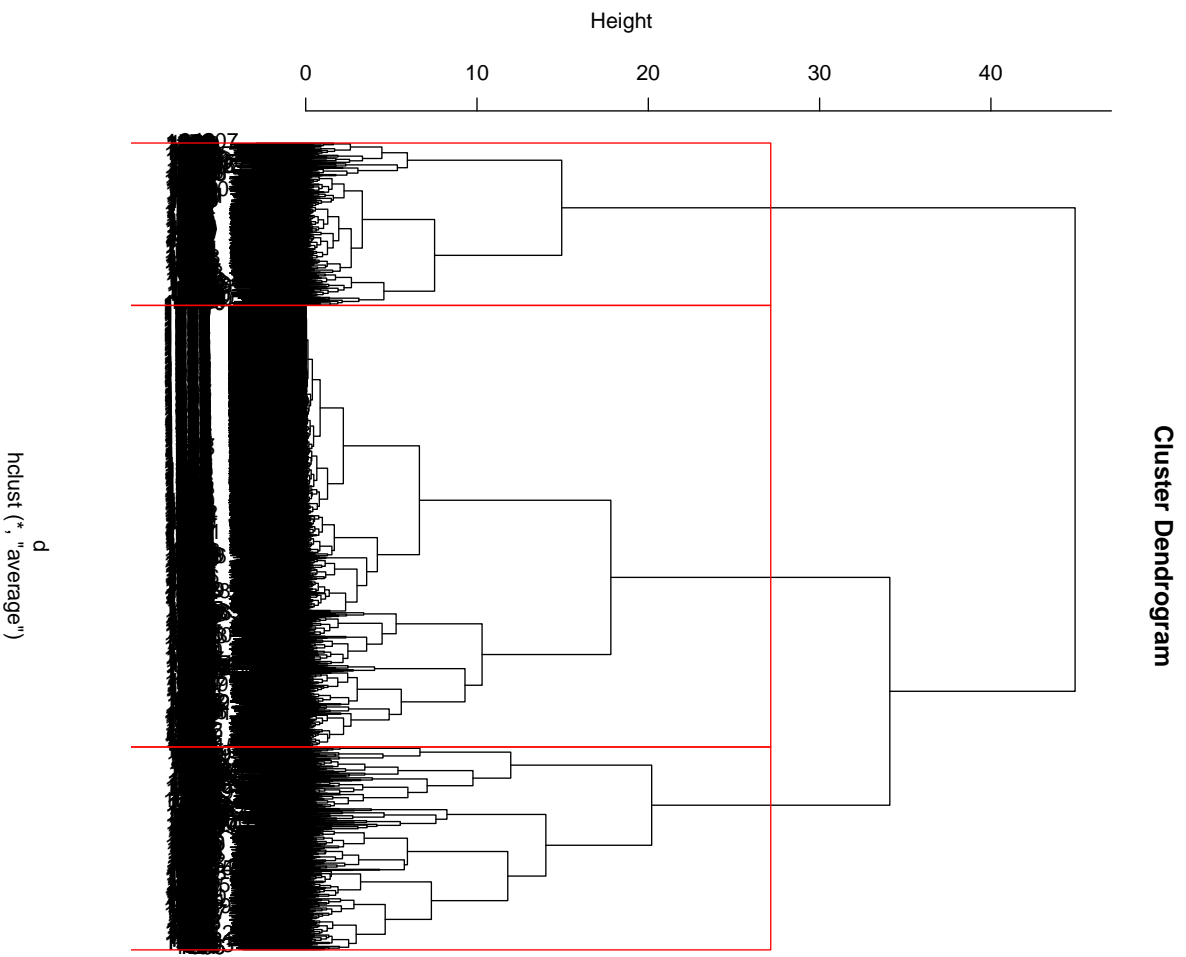}
\caption{Dendrogram obtained from hierarchical clustering applied to the first two KPCs, extracted through kernel (\ref{my kernel}) with $p=1/2$ and $s=\sigma_{1}$, revealed three clusters of GRBs.}\label{den}
\end{figure}
\clearpage
\begin{figure}
\centering
\includegraphics[width=1\textwidth]{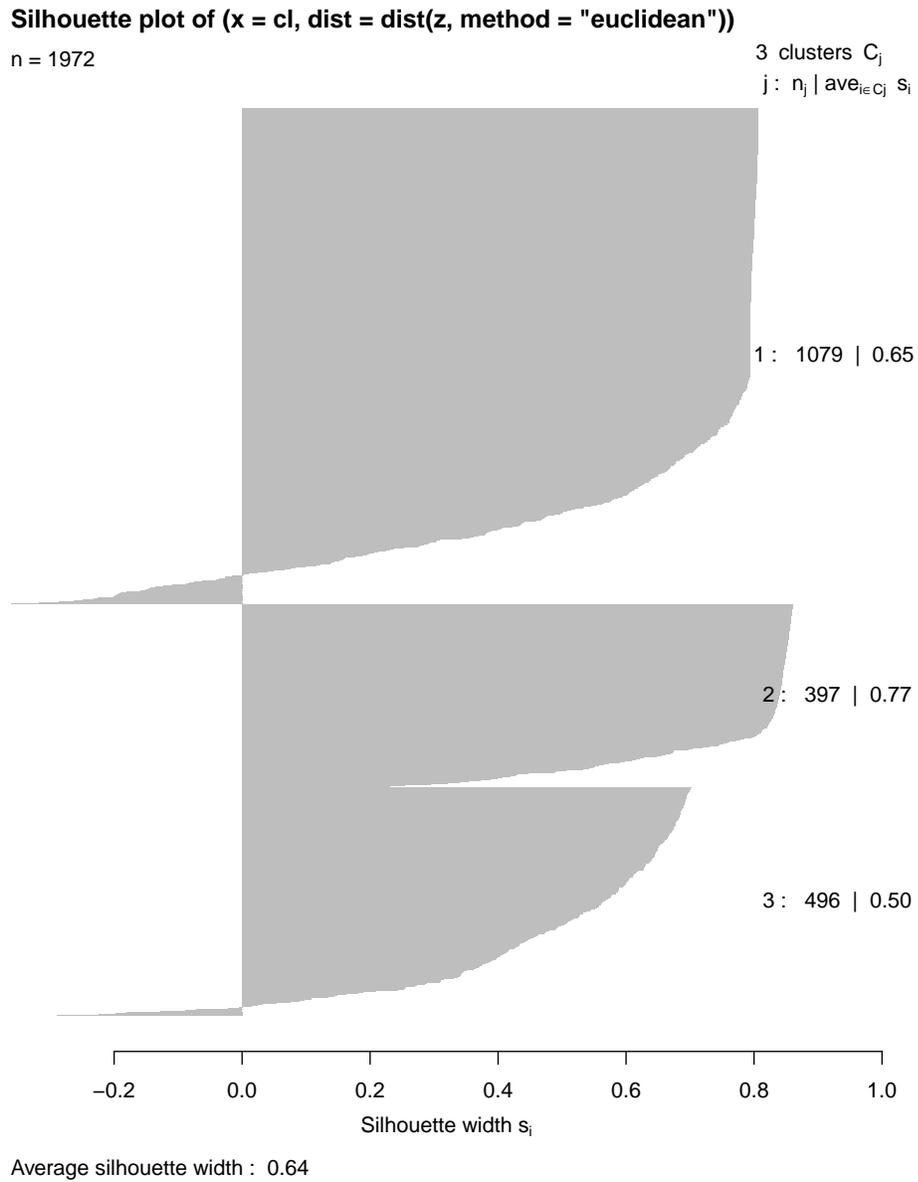}
\caption{Silhouette plot for three clusters of GRBs obtained from hierarchical clustering applied to the first two KPCs, extracted through kernel (\ref{my kernel}) with $p=1/2$ and $s=\sigma_{1}$.}\label{silplot}
\end{figure}
\clearpage
\begin{figure}
\centering
\includegraphics[width=1\textwidth]{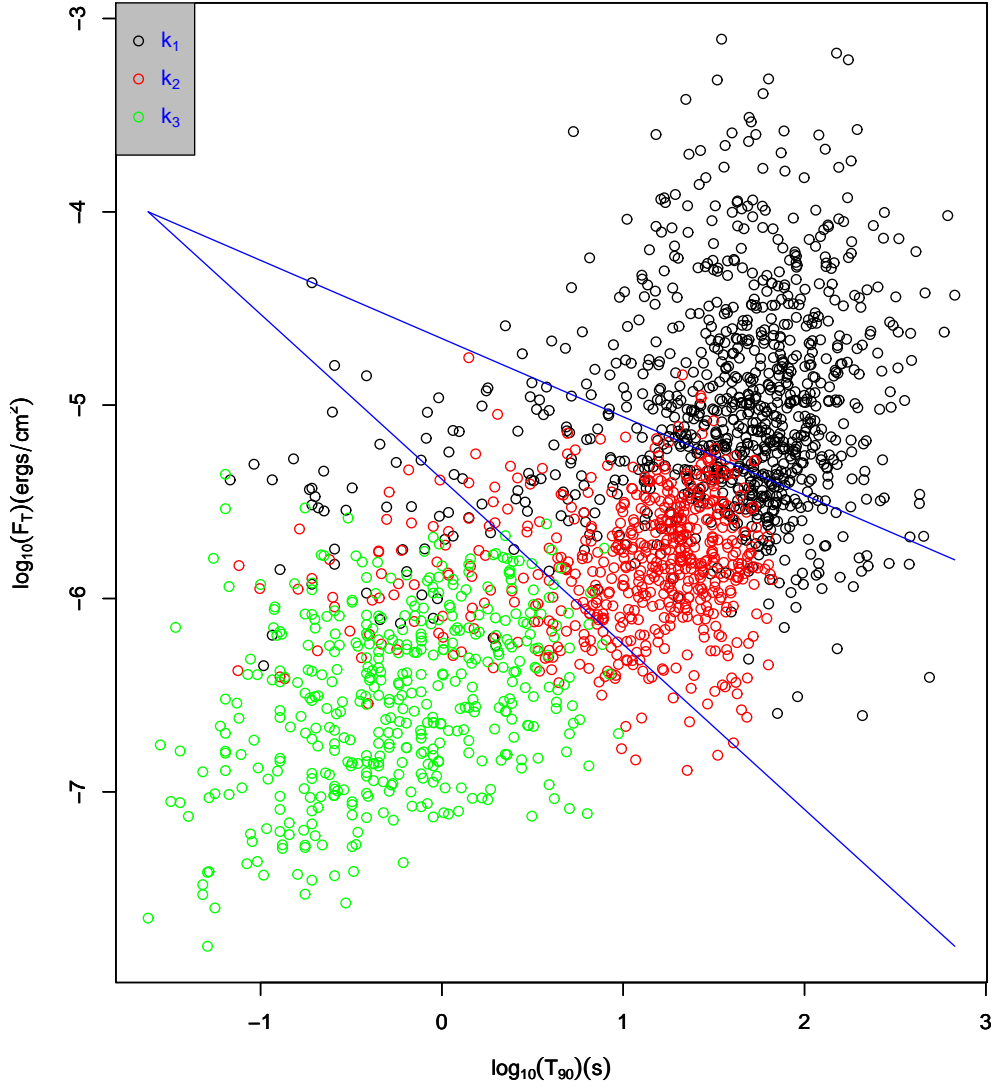}
\caption{$log_{10}(T_{90})$ (in s) vs. $log_{10}(F_{T})$ (in ergs cm$^{-2}$) plot for three clusters of GRBs. The solid lines represent $F_T=10^{-5.4}/T_{90}^{0.9}$ ergs cm$^{-2}$ (bottom) which separates short duration bursts, i.e. $k_3$ from long duration bursts and $F_T=10^{-4.6}/T_{90}^{0.4}$ ergs cm$^{-2}$ (top) which separates long duration bursts further into $k_1$ and $k_2$.}
\label{TimevsFluencewithlines}
\end{figure}
\clearpage
\begin{figure}
\centering
\includegraphics[width=1\textwidth]{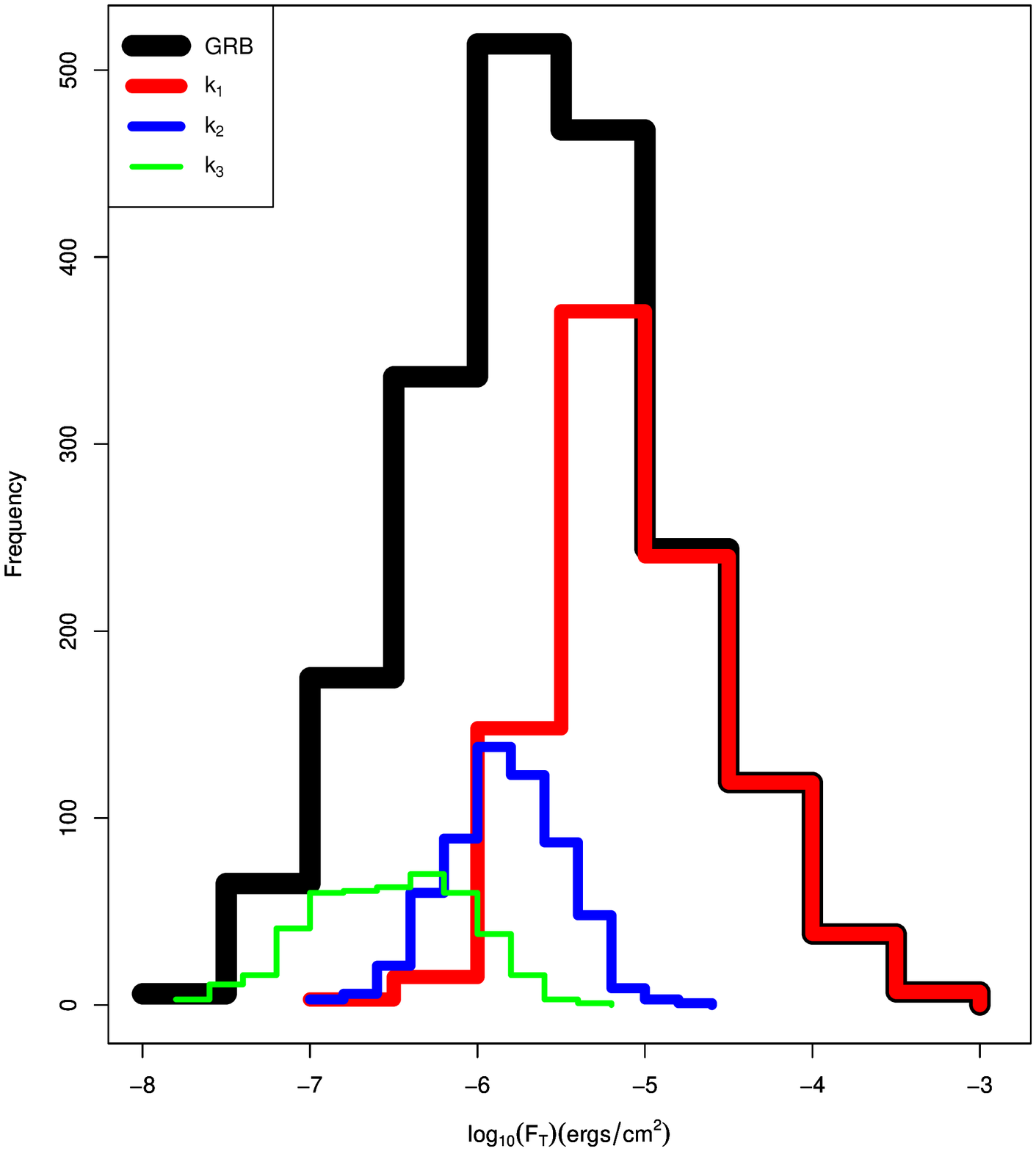}
\caption{Histograms of $log_{10}(F_T)$ for all the bursts (GRB) and three clusters ($k_1,k_2,k_3$).}\label{HistFt}
\end{figure}
\clearpage
\begin{figure}
\centering
\includegraphics[width=1\textwidth]{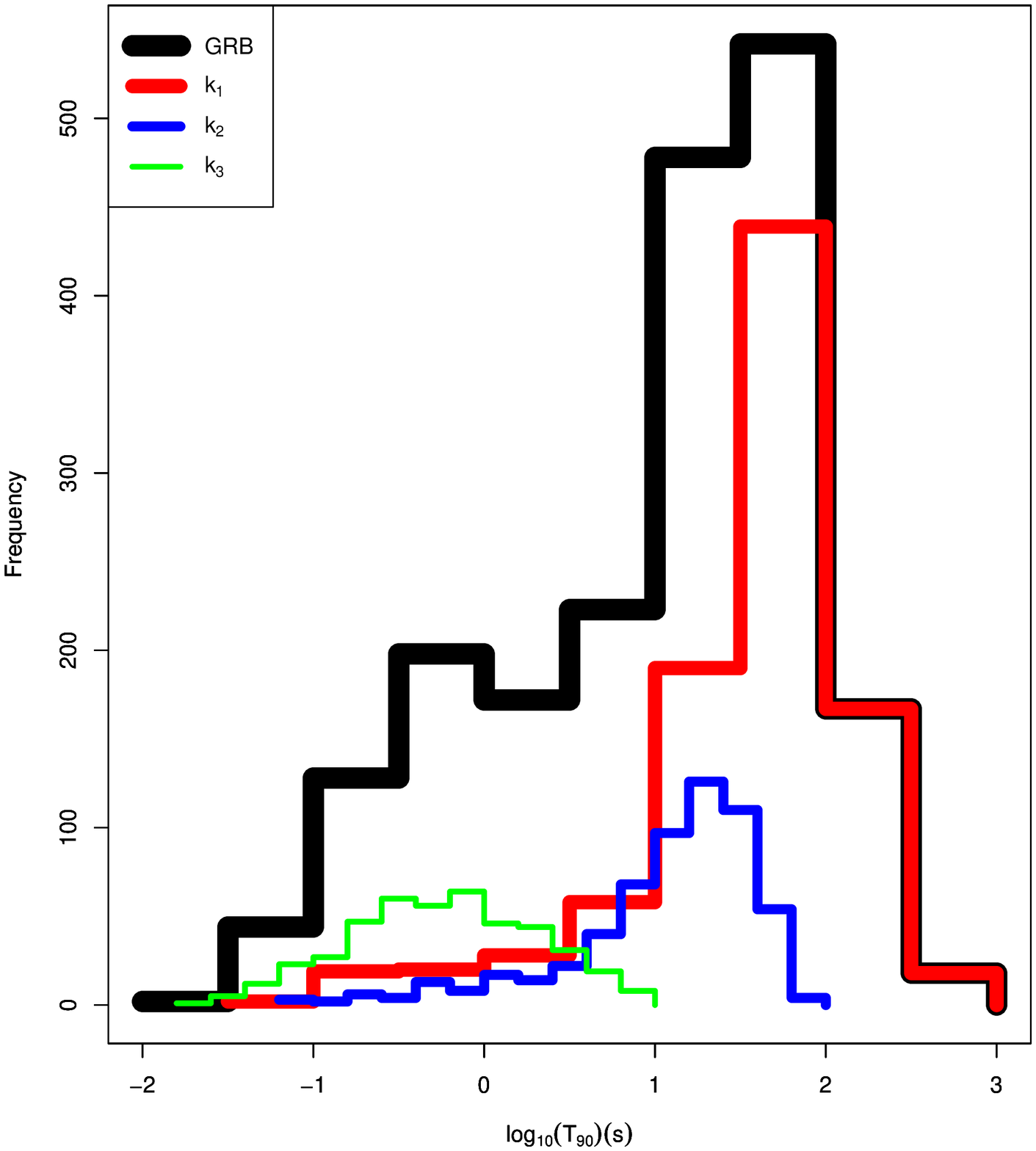}
\caption{Histograms of $log_{10}(T_{90})$ for all the bursts (GRB) and three clusters ($k_1,k_2,k_3$).}\label{HistT90}
\end{figure}
\clearpage
\begin{figure}
\centering
\includegraphics[width=1\textwidth]{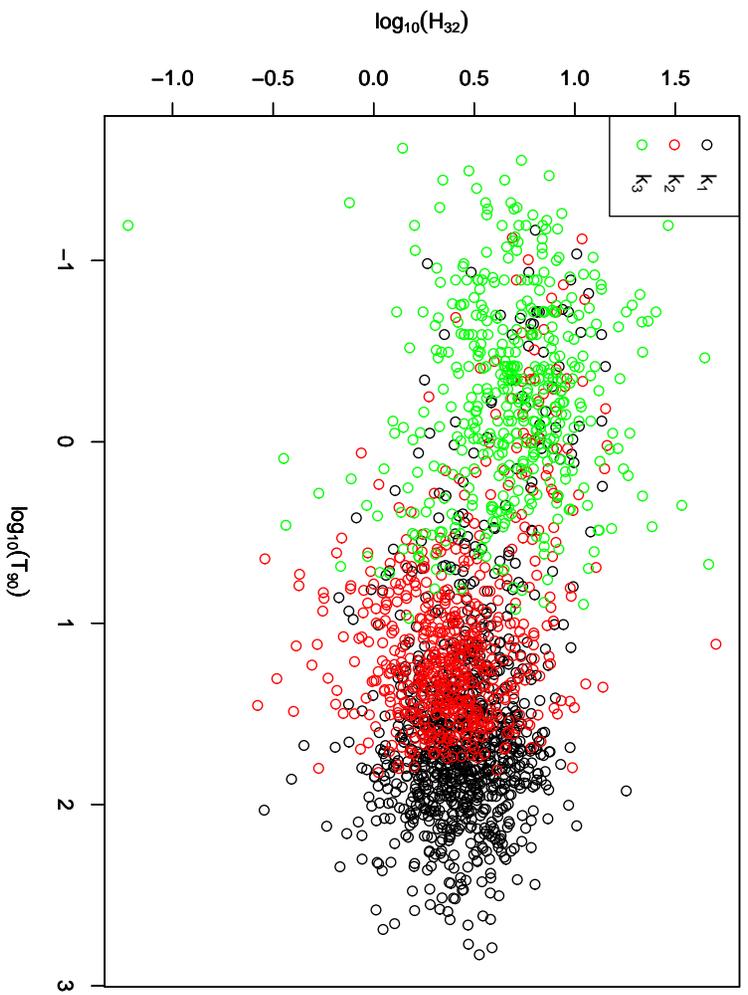}
\caption{Plot of $log_{10}(T_{90})$ vs. $log_{10}(H_{32})$ for three clusters.}\label{H32}
\end{figure}
\clearpage
\begin{figure}
\centering
\includegraphics[width=1\textwidth]{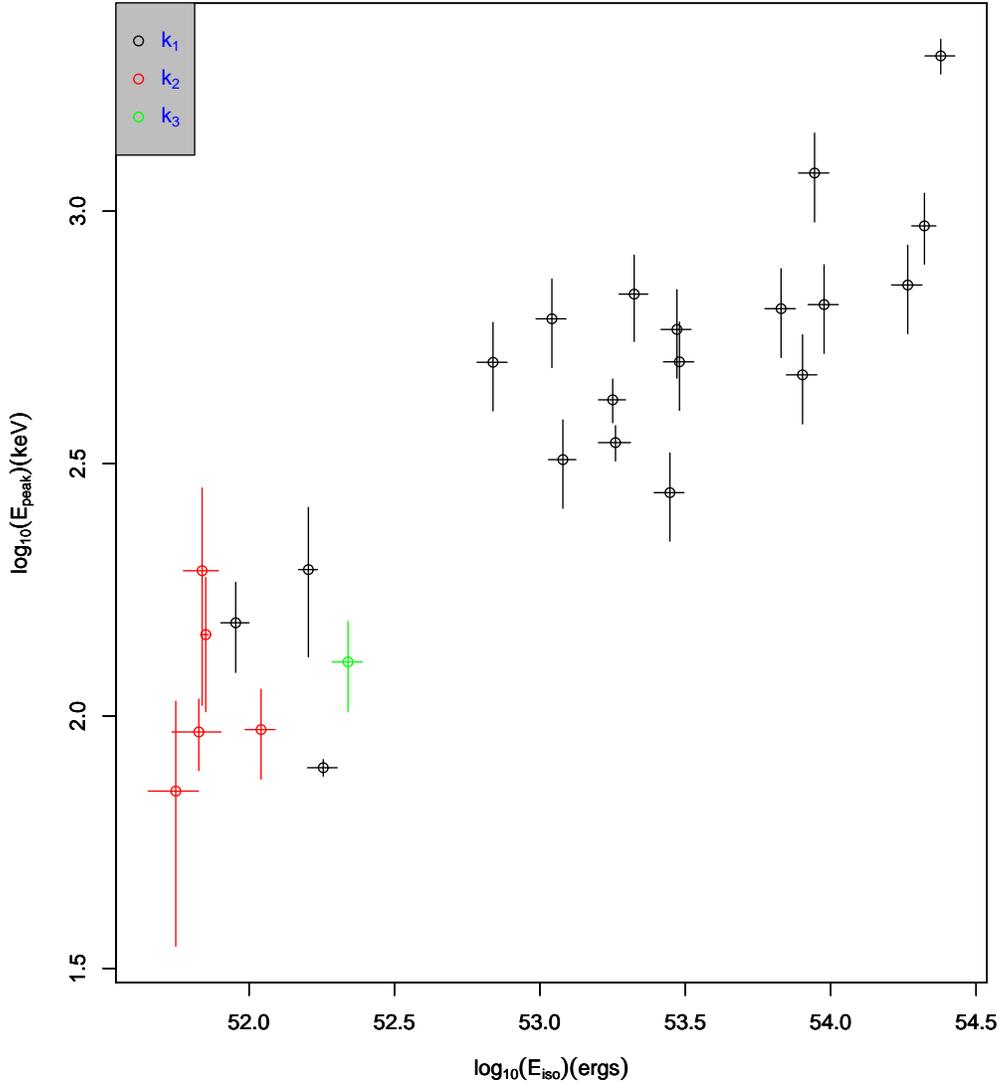}
\caption{log$_{10}(E_{peak})$ (in keV) vs.
log$_{10}(E_{iso}$) (in ergs) plot for 25 GRBs with well-defined spectral parameters
(Ghirlanda et al., 2004).
}\label{ClassificationGRB}
\end{figure}
\clearpage
\section{Acknowledgements}
The authors would like to show their gratitude to Professor Uttam Bandyopadhyay, Department of Statistics, University of Calcutta, India, for helping with the work. The authors thank the referee for constructive comments to improve the quality of the paper.


\begin{thebibliography}{99}
\bibitem{}
Balastegui, A., Ruiz-Lapuente, P., \& Canal, R. (2001). \textsl{Reclassification of gamma-ray bursts.} Mon. Not. R. Astron. Soc. \textbf{328}, 283--290.\\
\bibitem{}
Berg, C., Christensen, J. P. R. \& Ressel, P. (1984). \textsl{Harmonic Analysis on Semigroups,
Theory of Positive Definite and Related Functions.} \textbf{100}. New York: Springer. \\
\bibitem{}
Chattopadhyay, T., Misra, R., Chattopadhyay, A. K., \& Naskar, M. (2007). \textsl{Statistical evidence for three classes of gamma-ray bursts}. Astrophys. J. \textbf{667}, 1017--1023.\\
\bibitem{}
Dunn, JC (1974). \textsl{Well-separated clusters and optimal fuzzy partitions.}  J. Cybernetics. \textbf{4}, 95--104.\\
\bibitem{R}
Efron, B. \& Tibshirani, R. (1993). \textsl{An Introduction to the Bootstrap.} New York: Chapman \& Hall. \\
\bibitem{R}
Ghirlanda, G., Ghisellini, G. \& Lazzati, D. (2004). \textsl{The collimation-corrected gamma-ray burst eneregies correlate
with the peak energy pf their $\nu F_{\nu}$ spectrum}. Astrophys. J. \textbf{616}, 331--338\\
\bibitem{}
Hartigan, J. A. \&  Wong, M. A. (1979). \textsl{A K-means clustering algorithm.} Applied Statistics. \textbf{28}, 100--108. \\
\bibitem{}
Hofmann, T., Sch\"{o}lkopf, B. \& Smola, A. (2008). \textsl{Kernel methods in machine learning.} Ann. Statist. \textbf{36}, 1171--1220.\\
\bibitem{}
Ishida, E. E. O. \& Souza, R. S. de. (2013). \textsl{Kernel PCA for type Ia supernovae photometric
classification}. Mon. Not. R. Astron. Soc. \textbf{430}, 509--532.\\
\bibitem{}
Ishida, E. E. O. (2012). \textsl{Kernel PCA for supernovae photometric classification.} Proceedings of the International Astronomical Union. \textbf{10}, 683--684.\\
\bibitem{}
Kaufman, L. \& Rousseeuw, P.J. (1990). \textsl{Finding Groups in Data: An Introduction to Cluster Analysis.} Wiley, New York. \\
\bibitem{}
King, A., Olsson, E., \& Davies, M. B. (2007). \textsl{A new type of long gamma-ray burst.} Mon. Not. R. Astron. Soc. \textbf{374}, L34.\\
\bibitem{}
Kouveliotou, C., Meegan, C. A., Fishman, G. J., Bhat, N. P., Briggs, M. S., Koshut, T. M., Paciesas, W. S., \& Pendleton, G. N. (1993). \textsl{Identification of two classes of gamma-ray bursts.} Astrophys. J. \textbf{413}, L101.\\
\bibitem{}
Modak, S., Chattopadhyay, T. \& Chattopadhyay, A. K. (2016). \textsl{Two phase formation of massive elliptical galaxies : study
through cross-correlation including spatial effect.} http://arxiv.org/abs/1611.05213v1.\\
\bibitem{}
Paciesas, W.S., Meegan, C.A. \& Pendleton, G. N. et al. (1999). \textsl{The fourth batse gamma-ray burst catalog (revised)}. Astrophys. J. Suppl. \textbf{122}, 465--495.\\
\bibitem{R11}
Ripley, B. D. (1996). Pattern Recognition and Neural Networks. Cambridge: Cambridge University Press.\\
\bibitem{}
Rousseeuw, P. J. (1987). \textsl{Silhouettes: A graphical aid to the interpretation
and validation of cluster analysis}. Journal of Computational and Applied
Mathematics. \textbf{20}, 53--65.\\
\bibitem{}
Sch\"{o}lkopf, B. \& Smola, A. (2002). Learning with kernels: Support vector machines, regularization, optimization, and beyond. MIT Press.\\
\bibitem{}
Tibshirani, R., Walther, G. \& Hastie, T. (2001). \textsl{Estimating the number of clusters in a data set via the gap statistic}. J. Roy. Statist. Soc. Ser. B. \textbf{63}, 411--423.\\
\bibitem{}
Veres, P., Bagoly, Z., Horv$\acute{a}$th, I., M$\acute{e}$sz$\acute{a}$ros, A., Bal$\acute{a}$zs, L. G. (2010). \textsl{A Distinct Peak-flux Distribution of the Third Class of Gamma-ray Bursts: A Possible Signature of X-ray Flashes?} Astrophys. J. \textbf{725}, 1955–-1964.\\
\end{thebibliography}
\end{document}